\documentclass[aps,prc,twocolumn,groupedaddress,showpacs]{revtex4}
\newcommand \bea{\begin{eqnarray}}
\newcommand \eea{\end{eqnarray}}
\newcommand \beq{\begin{eqnarray}}
\newcommand \eeq{\end{eqnarray}}

\newcommand{\ve}[1]{\mbox{\boldmath $#1$}}

\usepackage{graphicx}
\begin{document}


\title{Long-wavelength spin- and spin-isospin correlations in nucleon matter}


\author{G. I. Lykasov}
\affiliation{JINR, Dubna 141980, Moscow Region, Russia}
\author{E. Olsson}
\affiliation{Department of Astronomy and Space Physics, Uppsala University, Box 515, 75120 Uppsala, Sweden}
\altaffiliation{NORDITA, Blegdamsvej 17, DK-2100 Copenhagen \O, Denmark}

\author{C. J. Pethick}
\affiliation{NORDITA, Blegdamsvej 17, DK-2100 Copenhagen \O, Denmark}

\date{\today}

\begin{abstract}

We analyse the long-wavelength response of a normal Fermi liquid using
Landau theory.  We consider contributions from intermediate states
containing one additional quasiparticle-quasihole pair as well as
those from states containing two or more additional
quasiparticle-quasihole pairs.  For the response of an operator
corresponding to a conserved quantity, we show that the behavior of
matrix elements to states with more than one additional
quasiparticle-quasihole pair at low excitation energies $\omega$
varies as $1/\omega$.  It is shown how rates of processes involving
transitions to two quasiparticle-quasihole states may be calculated in
terms of the collision integral in the Landau transport equation for
quasiparticles.

\end{abstract}

\pacs{05.30.Fk, 21.65.+f, 25.30.Pt, 26.50.+x, 26.60+c}

\maketitle


\section{Introduction}

In liquid helium 3, the prototypical Fermi liquid, interactions
between atoms are to a very good approximation invariant under independent rotations
in coordinate space and in spin space.  This fact, together with the
translational invariance of the system, makes for an economical
description of the properties of the system in terms of Landau's
theory of normal Fermi liquids.  The reason for this is that at long
wavelengths, the only excitations of importance for the low-frequency
response are those with an added quasiparticle-quasihole pair, and
collective modes such as zero sound, which are coherent superpositions
of single quasiparticle-quasihole pairs. An analysis of the behavior
of matrix elements of the long-wavelength components of the density
operator to single-pair and multipair states was given by Pines and
Nozi\`eres \cite{pines}, and our purpose in this paper is to extend
the analysis to more general responses.

From investigations of the response to probes more general than a
scalar potential \cite{leggett, bp}, it is clear that conservation
laws play a key role in determining the form of long-wavelength matrix
elements of an operator.  In nuclear matter, interactions are not
central, due to the large tensor contribution coming from exchange of
pions and to other terms that give rise to spin-orbit coupling.  As a
consequence, the total spin is not a conserved quantity and matrix
elements to states containing more than one quasiparticle-quasihole
pair (which we shall in future refer to simply as a pair) do not
vanish at long wavelengths.  A qualitative discussion of the effects
of noncentral interactions is given in Ref.\ \cite{op} and Landau
theory was extended to take into account noncentral interactions in
Ref.\
\cite{ohp}.

Multipair states play an important role in the context of neutrino
processes in dense matter.  The rates of a variety of weak interaction
processes can be expressed in terms of correlation functions for
nucleons or other constituents \cite{raffelt1}.  The reason for this
simplification is that, due to the interactions between leptons and
hadrons being due to weak and electromagnetic effects, rates may be
expressed as a convolution of a hadronic correlation function and a
leptonic one.  Because correlations among leptons are relatively weak,
the most difficult part of calculating weak interaction processes in
dense matter is the evaluation of the hadronic correlation functions.
A review of calculations of rates may be found in Ref.\
\cite{prakash}.  Processes such as bremsstrahlung of
neutrino-antineutrino pairs in nucleon-nucleon interactions and the
production of neutrinos and antineutrinos by the modified Urca process
are both intrinsically two quasiparticle, two quasihole processes, and
they are important when the corresponding single-pair process is
kinematically forbidden.  Bremsstrahlung by creation of a single
nucleon pair is kinematically forbidden, since the velocity of a
quasiparticle is always less than $c$, and the direct Urca process, the
single nucleon pair analog of the modified Urca process, is allowed in
dense matter only for certain compositions.  There are many
calculations of rate of scattering of neutrinos by nucleons that take
into account single-pair processes \cite{chris1,reddy} but,  as
Raffelt and collaborators have stressed
\cite{raffelt2}, multipair processes can be important quantitatively. 

The purpose of this paper is to express the
two-quasiparticle-quasihole contribution to the response in terms of
the collision integral in the Landau kinetic equation for
quasiparticles. Landau theory provides a clear separation between
long-wavelength degrees of freedom, which are treated explicitly, and
short-wavelength ones whose effects are included through the use of
parameters that include the effects of renormalization of matrix
elements of currents and interparticle interactions.  Another strength
of the Landau theory is that it brings out clearly the role played by
conservation laws.  For simplicity, we shall restrict ourselves to
normal systems and we shall not consider effects of nucleon pairing
and superfluidity.

\section{Basic formalism}

In Landau theory, one limits oneself to temperatures low compared with
the Fermi temperatures of the constituents and regards the elementary
excitations as being quasiparticles which may be characterized by an
effective mass $m^*_i$, where the index $i={\rm n}, {\rm p}$ denotes
neutrons and protons, respectively. We shall treat the nucleons as
nonrelativistic, which is a good first approximation.

If a system is subjected to a perturbation
\beq
H'=  {\cal O}_{\bf q} U_{\bf q}e^{-i\omega t} +{\rm Hermitean\
 conjugate},
\eeq
the Fourier transform of the linear response of the expectation value of
 the
operator ${\cal O}_{\bf q}$ at frequency $\omega$ is given for a system invariant under parity by \cite{pines}
\beq
<{\cal O}_{\bf q}>_{\omega} = -\chi(q, \omega) U_{\bf q},
\eeq
where the linear response function is defined by
\beq
\chi(q, \omega)=\sum_{jl} \frac{{\rm e}^{-E_l/T}}{Z} |({{\cal O}_{\bf q}^{\dagger}})_{jl}|^2 
\frac{2\omega_{jl}}{{\omega}_{jl}^2-(\omega+i\eta)^2}.
\label{response}
\eeq
Here $l$ denotes an initial state, $j$ denotes an intermediate state
 and
 $\omega_{jl}=E_j -
E_l$, where $E_j$ and $E_l$ are the energies of the states, $Z=\sum_j{\rm e}^{-E_j/T}$ is the partition function and $T$ is the temperature.
In this paper we shall only consider the response of the physical quantity which is coupled to the applied field.  
For systems with noncentral interactions, there are generally off-diagonal terms, such as the response of the spin in the $x$ direction when a magnetic field is applied in the $z$ direction.  Since such effects vanish in the limit of small $q$, they are generally of less importance than the diagonal response \cite{chris1}.  

A particularly useful quantity related to this is the dynamical structure factor, which is defined by
\bea
S_{\cal O}(q, \omega)=\frac{1}{Z} \sum_{jl} {\rm e }^{-E_l/T} |({{\cal
    O}_{\bf q }^{\dagger}})_{jl}|^2 \delta(\omega-\omega_{jl})\nonumber\\
 =   \frac{1}{\pi(1-{\rm e}^{-\omega/T})}{\rm Im } \chi(q, \omega).
\label{structure}
\eea

To make explicit calculations we shall use the language of diagrammatic 
many-body theory expressed in terms of fully renormalized nucleon
 propagator $G({\bf p}, \omega)$, where $\bf p$ is the particle momentum 
and $\omega$ its energy.  For simplicity we shall not write out spin and 
isospin labels except where necessary for clarity.  
We shall further write the propagator as the sum of a coherent part, which corresponds to a single quasiparticle or quasihole,  
and an incoherent part, which takes into account more complicated intermediate states:
\beq
G(p, \omega)= G^{\rm coh}(p, \omega) + G^{\rm inc}(p, \omega),
\eeq
where the coherent part has the form 
\beq
G^{\rm coh}(p, \omega)= \frac{z_p}{\omega -\epsilon_p +i\eta}.
\eeq
Here $\epsilon_p$ is the quasiparticle energy, measured with respect to the chemical potential for the species in question and $z_p$ is the wave function renormalization factor.

\begin{figure}
\includegraphics[width=0.4\textwidth,angle=0]{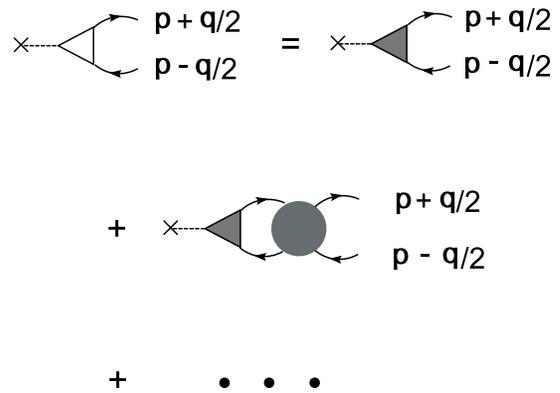}
\centering
\caption{Diagrams for the matrix element for a transition to a state with one quasiparticle-quasihole pair. The dashed line represents the interaction with the applied external field. The open circle corresponds to the sum of all connected diagrams, and the shaded ones to all connected diagrams irreducible with respect to a single quasiparticle-quasihole pair.}
\label{singlepair}
\end{figure}

\section{Single-pair processes}

To set the scene and establish notation, we consider transitions to states with a single quasiparticle-quasihole pair.  The matrix element for the process is indicated diagrammatically in Fig. \ref{singlepair} by the open circle. In Fig. \ref{singlepair} the single particle propagator represent only the coherent part. By the standard techniques of microscopic many-body theory \cite{nozieres}, this may be expressed in terms of diagrams irreducible with respect to states having a single quasiparticle-quasihole pair in the channel with total momentum $\bf q$ as shown in the figure, where the hatched vertices correspond to the irreducible diagams. Algebraically, it is given in a compact matrix notation by
\beq
{\cal O}_{\bf q}^{\rm sp}= \Lambda^{\cal O}z(1+\chi f)^{-1}\equiv
\lambda^{\cal O},
\label{lambdadef}
\eeq

where we suppress sums over momenta and particle species
Here $\Lambda$ is the irreducible vertex for interaction
with the external field,

\beq
\chi_{\bf p}( {\bf q}, \omega) =   \frac{ n_{\bf p+q/2}- n_{\bf p-q/2}}
{\omega-\epsilon_{\bf p+q/2}+ \epsilon_{\bf p-q/2}},
\eeq
and 
\beq
f_{\bf p p'}=z_pz_{p'} \Gamma^{\rm irr}({\bf p, p'})
\eeq
is the Landau quasiparticle interaction, $ \Gamma^{\rm irr}$ being the
two-particle vertex function irreducible in the sense described
above. We will give expressions for $\lambda^{\cal O}$  in the
Appendix. For the
long-wavelength, low-frequency response the only quasiparticles of
importance are those to the Fermi surface, and we denote $z_{p_{\rm
F}}$ by $z$. 
In the phenomenological theory, the quantity $g^{\cal O}\equiv z\Lambda^{\cal O}$ corresponds to an effective ``charge'' of a quasiparticle for the particular probe in question; for example in the case of a scalar probe, it is the total particle number carried by a quasiparticle and for a magnetic field it corresponds to the magnetic moment of the quasiparticle.
For simplicity we have not written spin and isospin indices explicitly.

The contribution to the imaginary part of the response function is therefore
\bea
{\rm Im} \chi^{\rm sp}&=& \pi \sum_{\bf p, spin}  |{\cal O}_{\bf q}^{\rm sp}|^2 (n_{\bf p+q/2}- n_{\bf p-q/2}) \times \nonumber \\
&&\delta(\omega-\epsilon_{\bf p+q/2}+ \epsilon_{\bf p-q/2} )  
\eea

\begin{figure}
\includegraphics[width=0.4\textwidth,angle=0]{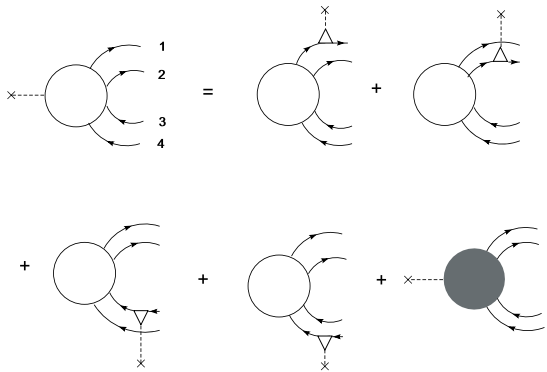}
\centering
\caption{Diagrams for the matrix element for a transition to a state
  with two quasiparticle-quasihole pairs. As in Figure 1, open circles denote the sum of all connected diagrams, and here the hatched circle corresponds to the sum of all diagrams that do not have the external field attached to one of the external lines.}
\label{twopair}
\end{figure}

\section{Two-pair processes}

We turn now to intermediate states with two quasiparticles and two quasiholes.
The matrix element for the process is illustrated in Fig. \ref{twopair}. Of particular interest to us are contributions in which the external field interacts directly with just one of the quasiparticles or quasiholes in the intermediate state.  These are represented by the first four diagrams on the right hand side of the equation in Fig. \ref{twopair}, and they are particularly important for the response to operators which do not correspond to conserved quantities.  As we shall see, they lead to divergences of matrix elements for low $\omega$ analogous to those found for bremsstrahlung.  The corresponding effects in low-order perturbation theory have previously been drawn attention to in Ref.\ \cite{raffelt2}. The final term in Fig. \ref{twopair} represents the remainder.  We denote the quasiparticles and quasiholes in the intermediate state by the indices $i=1,\ldots,4$, as indicated. In this paper we shall consider only operators which do not convert neutrons into protons, or vice versa, that is they couple only to the third component of the isospin.  When the operator can change the isospin, which is the case for the response function of interest for the modified Urca process, the situation is more complicated, and we shall comment on this point in the Conclusion.

 For small $\bf q$, the contribution to the on-shell matrix element coming from the first term may be evaluated straightforwardly, and it is given by
\bea
{\cal O}_{\bf q}^{\rm tp}= -z^2\Gamma \frac{1}{\omega - \epsilon_{{\bf p}_1} + 
\epsilon_{{\bf p}_1-{\bf q}}}\lambda^{\cal O}_1
\nonumber\\
\approx -{\cal A}\frac{\lambda^{\cal O}_1 }{\omega -{\bf v_{\rm _1}\cdot q}},
\eea
where 
\beq
{\cal A}= z^2 \Gamma
\eeq
is the scattering matrix for two-quasiparticle scattering.
Therefore the contribution to the  matrix element from the first four terms is
\bea
{\cal O}_{\bf q}^{\rm tp}= -{\cal A}\frac{1}{\omega}\left[\xi^{\cal O}_1 +
\xi^{\cal O}_2-\xi^{\cal O}_3-\xi^{\cal O}_4\right],
\label{metp}
\eea
where
\beq
\xi^{\cal O}_i=\frac{\lambda^{\cal O}_i}{1- {\bf v_{\rm_i}\cdot q}/\omega}. 
\label{Eq:xi}
\eeq
The imaginary part of the two-pair contribution to the response function is therefore
\bea
{\rm Im} \chi^{\rm tp} =\frac{\pi}{4} \sum_{1234} \left\{ |{\cal O}_{\bf q}^{\rm tp}|^2\times\right.\nonumber\\
\left[ n_1 n_2 (1 - n_3 )(1 - n_4 )-(1- n_1 )(1- n_2 ) n_3 n_4 \right] \times\nonumber\\
\left.\delta(\omega+\epsilon_1+ \epsilon_2-\epsilon_3 -\epsilon_4 )\delta( {\bf q+p_1+p_2-p_3-p_4})\right\}.\nonumber\\
\eea
where sums over quasiparticle states include sums over spins and isospins and the factor of $1/4$ is necessary to avoid overcounting.
Inserting the expression (\ref{metp}), one finds \cite{traces}
\bea
 {\rm Im} \chi^{\rm tp}=\pi \!\sum_{1234}\!\! \left\{ |{\cal A}|^2 \left(\frac{1}{\omega^2}\xi^{\cal O}_1\right)^{\scriptstyle *} \!\!\!    
\left( \xi^{\cal O}_1+\xi^{\cal O}_2-\xi^{\cal O}_3-\xi^{\cal O}_4)\right)\right.\times\nonumber\\
\left[ n_1n_2(1-n_3)(1-n_4)-(1-n_1)(1-n_2)n_3n_4 \right]\times\nonumber\\ 
\left.\delta(\omega+\epsilon_1+\epsilon_2-\epsilon_3 -\epsilon_4 )
\delta({\bf q+p_1+p_2-p_3-p_4})\frac{}{}\right\}\nonumber\\
\label{chitp}
\eea
Let us now consider the limit $\omega >> v_{\rm F}q$.
For external fields that are independent of the particle
momentum, the quantities $\lambda$ are independent of the direction of the quasiparticle
momentum.  For long-wavelength probes, the  $\bf q$ in the momentum
conservation condition may be ignored to leading order, and therefore Eq. (\ref{chitp}) becomes   
\bea
 {\rm Im} \chi^{\rm tp}= \frac{\pi}{\omega^2} \sum_{1234}\left\{  |{\cal A}|^2 \xi^{\cal O}_1 (\xi^{\cal O}_1 +\xi^{\cal O}_2 -\xi^{\cal O}_3-\xi^{\cal O}_4)\right.\times\nonumber\\
\left[ n_1n_2(1-n_3)(1-n_4)-(1-n_1)(1-n_2)n_3n_4 \right]\times\nonumber\\ 
\left.\delta(\omega+\epsilon_1+ \epsilon_2-\epsilon_3 -\epsilon_4 )\delta( {\bf p_1+p_2-p_3-p_4})\right\}\nonumber\\
\label{chitp2}
\eea
For convenience, we adopt the usual convention that the phase of single-particle states is chosen so that $\lambda^{\cal O}$ is real.

It is instructive to rewrite this expression to show that the quantity that enters is very closely related to quasiparticle relaxation time.  The collision
 term for two-quasiparticle scattering is given by 
\bea
\frac{\partial n_1}{\partial t}=-\sum_{234}\left\{  |{\cal A}|^2\right.\times \nonumber\\
\left[ n_1n_2(1-n_3)(1-n_4)-(1-n_1)(1-n_2)n_3n_4 \right]\times\nonumber\\ 
\left.\delta(\omega+\epsilon_1+ \epsilon_2-\epsilon_3 -\epsilon_4 )\delta( {\bf p_1+p_2-p_3-p_4})\right\},
\eea
where in this equation the distribution function is not necessarily the equilibrium one.  On linearizing this equation about the equilibrium distribution for quasiparticles with momentum $\bf p_1$, one finds

\beq
\frac{\partial n_1}{\partial t}=-\frac{\delta n_1}{\tau_1}
\eeq
where
\beq
\frac{1}{\tau_1}=2\pi\sum_{234}\left\{  |{\cal A}|^2 
[n_2(1-n_3)(1-n_4)+(1-n_2)n_3n_4 ]\right.\!\!\times\nonumber\\ 
\left.\delta(\omega+\epsilon_1+ \epsilon_2-\epsilon_3 -\epsilon_4 )\delta( {\bf p_1+p_2-p_3-p_4})\right\}.
\label{collisiontime}
\eeq
We now rewrite Eq. (\ref{chitp2}) so that the combinations of distribution functions is similar to that in Eq.\ (\ref{collisiontime}).
By virtue of the relationship
\beq
n(\epsilon)= {\rm e}^{-\epsilon/T}[1-n(\epsilon)]
\eeq
we may write
\bea
[n_1n_2(1-n_3)(1-n_4)-(1-n_1)(1-n_2)n_3n_4 ] \times\nonumber\\
\delta(\omega+\epsilon_1+ \epsilon_2-\epsilon_3 -\epsilon_4 )=\nonumber\\
\left[ n_1-n(\epsilon_1+\omega)][n_2(1-n_3)(1-n_4)+(1-n_2)n_3n_4 \right]\times\nonumber\\
\delta(\omega+\epsilon_1+ \epsilon_2-\epsilon_3 -\epsilon_4 ),\nonumber\\
\eea
from which it follows that we may rewrite
Eq.\ (\ref{chitp2}) as
\bea
 {\rm Im} \chi^{\rm tp}=\frac{1}{2\omega^2}\sum_1   
(\lambda^{\cal O}_1)^2  \frac{[n_1-n(\epsilon_1+\omega)]}{\tau^{\cal O}_1(\epsilon_1+\omega)}
\eea
where 
\bea
\frac{1}{\tau^{\cal O}_1(\epsilon_1 + \omega)}=2\pi\sum_{234}\left\{  |{\cal A}|^2 \left[1+\frac{     
\lambda^{\cal O}_2 -\lambda^{\cal O}_3 -
\lambda^{\cal O}_4}{ \lambda^{\cal O}_1 }\right]\right.\times\nonumber\\ 
\left[n_2(1-n_3)(1-n_4)+(1-n_2)n_3n_4 \right]\times\nonumber\\ 
\left.\frac{}{}\!\!\delta(\omega+\epsilon_1+ \epsilon_2-\epsilon_3 -\epsilon_4 )\delta( {\bf q+p_1+p_2-p_3-p_4})\right\}\nonumber\\
\label{reltime}
\eea
is the relaxation rate for the quantity 
$\lambda^{\cal O}$ for a system close to equilibrium. This gives the
collision rate weighted by the change in $\lambda$ of all the
quasiparticles in a collision. 

Depending on the properties of the operator $\cal O$, the matrix elements to two pair states can have different sorts of behavior. Let us begin by considering the limit $\omega\gg v_{\rm F}q$, where $v_{\rm F}$ is a typical Fermi velocity.  The matrix element then 
becomes proportional to $[\lambda^{\cal O}({\bf p}_1)+\lambda^{\cal O}({\bf p}_2)-\lambda^{\cal O}({\bf p}_3)-\lambda^{\cal O}({\bf p}_4)]/\omega$.  
If the operator $\cal O$ is the particle density, $\lambda^{\cal O}$ becomes a constant at for $\omega \gg v_{\rm F}q$.  Since the interaction conserves particle number, it also conserves the number of quasiparticles and therefore in this limit, $\lambda^{\cal O}({\bf p}_1)+\lambda^{\cal O}({\bf p}_2)-\lambda^{\cal O}({\bf p}_3)-\lambda^{\cal O}({\bf p}_4)$ vanishes.  It the operator is that for a component of the spin, a similar argument will apply if the interactions between components are central, that is their spin structure of the form of the unit operator or of the spin-exchange operator.  However, when there are non-central interactions, e.g. spin-orbit or tensor forces, the combination of $\lambda$'s will not generally vanish.  For example, two quasiparticles with spin up can scatter to states with two quasiparticles with spin down, in which case the matrix element to a two-pair state will diverge as $1/\omega$.  For the operator for the number density of particles, the leading contribution to the matrix element for a transition to a two-pair state will have a factor $v_{\rm F}q/\omega^2$. 
The matrix element will then vanish at least as rapidly as $q$ for $q\rightarrow 0$, in accordance with the general arguments in Ref.\ \cite{pines}.  However, if the corresponding current is conserved in the long-wavelength limit, as it will be if the system is translationally invariant, the term proportional to $q$ will also vanish, and the leading term will be or order $q^2$. For a non-relativistic system with a single component, the particle current is proportional to the total momentum, which is conserved for a translationally invariant system.  However, for a multicomponent system, the total particle current is proportional to the total momentum if all components have the same mass, and therefore for arbitrary masses of the components, the two-pair matrix element will be proportional to $q$.  For a system with components that do not all have the same mass, it may be seen from this argument that matrix elements of the {\it mass density} operator to two-pair states will vary as $q^2$ for small $q$, since the total mass current is conserved.

\section{Two-pair contributions to structure functions}

\begin{figure}
\includegraphics[width=0.45\textwidth,angle=0]{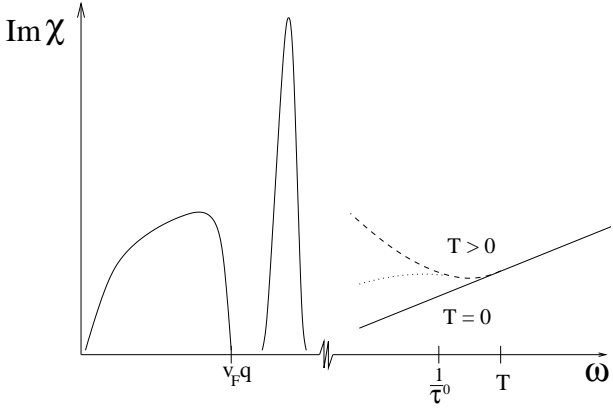}
\centering
\caption{A schematic picture of the behavior of the response for small
$q$. The bump at $\omega < v_{\rm F}q$ is the contribution from
single-pair states, the spike is due to the collective mode. The straight line represents the behavior of the two-pair
response for zero temperature, while the dashed line represents the
two-pair response for nonzero temperature, neglecting the multiple
scattering effect, which is included in the dotted line.}
\label{schematic}
\end{figure}

We now evaluate the sums over momenta in Eq.\ (\ref{reltime}).  For definiteness, we shall assume that quasiparticles 1 and 3 are the same species $i$.  Since collisions conserve the numbers of neutrons and of protons, this implies that quasiparticles 2 and 4 are the same, but not necessarily the same as 1 and 3.  
All quasiparticles participating in a collision at low excitation energies lie close to the respective Fermi surfaces, so the collision geometry may be specified
in terms of the angle $\theta$ between $\bf p_1$ and $\bf p_2$, and $\phi$, the angle between the plane containing $\bf p_1$ and $\bf p_2$ and the plane containing
$\bf p_3$ and $\bf p_4$. We use the result \cite{bp, anderson}
\bea
\sum_{{\bf p}_2 {\bf p}_3 {\bf p}_4}\delta^{(3)}({\bf p}_1 + {\bf p}_2 - {\bf p}_3 - {\bf p}_4)= \nonumber\\
=\frac{m^*_i(m^*_j)^2}{(2\pi)^6}\int\beta_{ij}\sin \theta d\theta d\phi d\phi_2 d\epsilon_2 d\epsilon_3 d\epsilon_4,
\label{def:relat}
\eea 
where $\phi_2$ is the azimuthal angle of $\bf p_2$ with respect to $\bf p_1$ and 
\beq
\beta_{ij}=\frac{p_j}{(p_i^2 + p_j^2+2p_i p_j \cos \theta)^{1/2} }.
\eeq
Provided the $\lambda_i$ may be regarded as independent of the magnitude
of the quasiparticles momentum, which is a good approximation for the
response considered in this article, the integrals over energies and over momenta decouple, and from Eq.\ (\ref{chitp2}) we find that

\bea
{\rm Im} \chi^{\rm tp}= \frac{1}{\omega^2} \sum_{ij}\left\{ \frac{(m^*_i m^*_j)^2 }{(2\pi)^7}\right.\times\nonumber \\ 
\int d\Omega d\phi_2 \gamma_{ij} \sum_{\rm spins} |{\cal A}^{ij}|^2\xi^{\cal O}_1 (\xi^{\cal O}_1 +\xi^{\cal O}_2 -\xi^{\cal O}_3-\xi^{\cal O}_4)\times\nonumber \\
\int d\epsilon_1 d\epsilon_2 d\epsilon_3 d\epsilon_4 \delta(\omega+\epsilon_1+ \epsilon_2-\epsilon_3 -\epsilon_4 )\times\nonumber \\
\left.\frac{}{}\left[ n_1n_2(1-n_3)(1-n_4)-(1-n_1)(1-n_2)n_3n_4 \right]\right\} ,\nonumber\\
\label{chitp3}
\eea
where 
\beq
\gamma_{ij}= p_i \beta_{ij}=\frac{p_i p_j}{(p_i^2 + p_j^2+2p_i p_j \cos \theta)^{1/2} },
\eeq
is symmetric: $\gamma_{ij}=\gamma_{ji}$.
The integral over energies is \cite[p. 114]{bp}
\bea
\int d\epsilon_1 d\epsilon_2 d\epsilon_3 d\epsilon_4 \delta(\omega+\epsilon_1+ \epsilon_2-\epsilon_3 -\epsilon_4 )\times\nonumber\\
\left[ n_1n_2(1-n_3)(1-n_4)-(1-n_1)(1-n_2)n_3n_4 \right]=\nonumber\\
\frac{\omega[\omega^2 + (2\pi T)^2]}{6},\nonumber\\
\eea
and therefore we find
\bea
{\rm Im} \chi^{\rm tp}=  \frac{[\omega^2+(2\pi T)^2]}{6 \omega} \sum_{ij}\left\{\frac{(m^*_i m^*_j)^2 }{(2\pi)^7}\times\right.\nonumber \\\left.\int d\Omega d\phi_2 \gamma_{ij} \sum_{\rm spins}|{\cal A}^{ij}|^2
\xi^{\cal O}_1 (\xi^{\cal O}_{1i} +\xi^{\cal O}_{2j} -\xi^{\cal O}_{3i}-\xi^{\cal O}_{4j})\right\}.\nonumber \\
\label{chitp4}
\eea
which for $\omega >> v_{\rm F}q$ reduces to, 
\bea
{\rm Im} \chi^{\rm tp}= \frac{[\omega^2+(2\pi T)^2]}{6\omega} 
\sum_{ij}\left\{\frac{(m^*_i m^*_j)^2 }{(2\pi)^7}\right.\times\nonumber \\\left.\int 
d\Omega d\phi_2 \gamma_{ij} \sum_{\rm spins}|{\cal A}^{ij}|^2
\lambda^{\cal O}_1     
(\lambda^{\cal O}_{1i} +\lambda^{\cal O}_{2j} -\lambda^{\cal O}_{3i}-\lambda^{\cal O}_{4j})\right\},\nonumber \\
\label{chitp5}
\eea
where from Eq. (\ref{Eq:xi}) one can see that $\xi \rightarrow \lambda$ in this regime.
This result shows that at zero temperature, ${\rm Im} \chi^{\rm tp}$
is proportional to $\omega$ and independent of $q$.  This behavior
should be contrasted with that of the single-pair contribution, which
varies as $\omega/q$ for small $\omega/v_{\rm F}q$.  At nonzero
temperature, ${\rm Im} \chi^{\rm tp} \propto T^2/\omega$. The
divergence at low $\omega$ is not physical, and is a consequence of
our having neglected the effects of real processes on the matrix
element to two-pair states. When such effects are included, the
divergence will be cut off.  This is an example of the
Landau-Pomeranchuk-Migdal effect \cite{lpm}. As pointed out in Ref.\
\cite{raffelt2} in the context of neutrino scattering, these will
become important when $\omega$ becomes comparable to the quasiparticle
lifetime (Eq. (\ref{reltime})) which, by manipulations similar to those used
for ${\rm Im} \chi^{\rm tp}$ may be written in the form

\bea
\frac{1}{\tau^{\cal O}_i(\epsilon_1 + \omega)}=2\pi  \frac{(\omega+\epsilon_1)^2+(\pi T)^2}{2}\sum_{j}\left\{\frac{ m^*_i (m^*_j)^2}{(2\pi)^6}\right.\times \nonumber\\
\left.\int d\Omega d\phi_2\beta_{ij}{\sum_{\rm spins}}^\prime |{\cal A}^{ij}|^2  
\left[1+\frac{\lambda^{\cal O}_{2j} -\lambda^{\cal O}_{3i}-\lambda^{\cal O}_{4j}}{ 
\lambda^{\cal O}_{1i} }\right]\right\},
\label{reltime1}
\eea
where the prime on the $\sum_{\rm spins}'$ indicates that the sum over spin for quasiparticle 1 is excluded.
In Fig.\ \ref{schematic} we show a schematic plot of the behavior of  
 ${\rm Im} \chi$ for small $q$.

The quasiparticle scattering amplitude has the following form for
neutron matter (\cite{friman,schwenk1}) (for quasiparticles on the
Fermi surface)
\bea
{\cal A}_{\ve{\scriptscriptstyle{\sigma \sigma}}'} ({\bf q},{\bf
  q}',{\bf P})= {\cal A}_{\rm scalar}+{\cal A}_{\rm  spin\ exch.}\ve{\sigma}\cdot\ve{\sigma}'+\nonumber\\{\cal A}_{\rm  spin-orbit}i(\ve{\sigma}+\ve{\sigma}')\cdot\hat{\bf  q}\times\hat{\bf q}'+
{\cal A}_{\rm tensor}S(\hat{\bf q})+\nonumber\\{\cal  A}_{\rm exch.\
  tensor}S(\hat{\bf q}')+{\cal A}_{\rm cm\  tensor}S(\hat{\bf P})+\nonumber\\
{\cal A}_{\rm diff.\ vector}i(\ve{\sigma}-\ve{\sigma}')\cdot\hat{\bf  q}\times\hat{\bf P}+\nonumber \\{\cal A}_{\rm cross\ vector} (\ve{\sigma}\times\ve{\sigma}')\cdot(\hat{\bf  q}'\times\hat{\bf P}), \nonumber \\
\eea
where \cite{tensoroperator}  
\beq
S(\hat{\bf q}) = 3 \ve{\sigma}\cdot\hat{\bf  q} \ve{\sigma}'\cdot\hat{\bf  q} -
\ve{\sigma}\cdot\ve{\sigma}',
\eeq
and $S(\hat{\bf q}')$ and $S(\hat{\bf P})$ have the same structure,
with $\hat{\bf q}$ replaced by $\hat{\bf q}'$ respective $\hat{\bf
  P}$. 
where ${\bf q} = {\bf p}_1 - {\bf p}_3$, ${\bf q}' = {\bf p}_1 - {\bf
  p}_4$ and ${\bf P} = {\bf p}_1 + {\bf p}_2={\bf p}_3 + {\bf
  p}_4$. 
The amplitudes $\cal A$ are functions of the scalar invariants that can be made from the vectors $\bf q, q'$  and $P$.
For particles on the Fermi surface ${\bf q}$, ${\bf q}'$,
${\bf P}$ are orthogonal to each other, i. e. ${\bf q}\cdot{\bf
  q}'=0$, ${\bf q}\cdot{\bf P}=0$ and ${\bf q}'\cdot{\bf P}=0$. For
asymmetric nuclear matter, the Fermi surfaces are different, and in
the case when 1 and 3 are of species $i$ and 2 and 4 are of species
$j$, this means that ${\bf q}$ is orthogonal to both ${\bf q}'$ and
${\bf P}$, but ${\bf q}'$ is not orthogonal to ${\bf P}$, which
leads to the following extra terms in the quasiparticle scattering amplitude \cite{friman}:
one similar to the spin-orbit interaction:
\beq
{\cal A}^{ij}_{q'P\rm \ so}{\hat{\bf q}}'\cdot {\hat{\bf P}}i(\ve{\sigma}+\ve{\sigma}')\cdot\hat{\bf q}\times\hat{\bf P},
\eeq
one similar to the difference vector one
\beq
{\cal A}^{ij}_{q'P\ \rm diff}{\hat{\bf q}}'\cdot {\hat{\bf P}}i(\ve{\sigma}-\ve{\sigma}')\cdot\hat{\bf q}\times\hat{\bf P}
\eeq
and one similar to the quadratic spin-orbit
\bea
{\cal A}^{ij}_{q'P\ \rm quadr.\ so}{\hat{\bf q}}'\cdot {\hat{\bf P}} \left(\ve{\sigma}\cdot{\hat{\bf q}}'\ve{\sigma}'\cdot{\hat{\bf P}} +\right.\nonumber \\\ve{\sigma}\cdot{\hat{\bf P}} \ve{\sigma}'\cdot{\hat{\bf q}}'-\frac{2}{3}\left.{\hat{\bf q}}'\cdot{\hat{\bf P}} \ve{\sigma}\cdot\ve{\sigma}'\right).
\eea
leading to the following scattering amplitude:
\bea
{\cal A}^{ij}_{\ve{\scriptscriptstyle{\sigma \sigma}}'} ({\bf q},{\bf
  q}',{\bf P})= {\cal A}^{ij}_{\rm scalar}+{\cal A}^{ij}_{\rm
  spin\ exch.}\ve{\sigma}\cdot\ve{\sigma}'+\nonumber \\{\cal A}^{ij}_{\rm  spin-orbit}i(\ve{\sigma}+\ve{\sigma}')\cdot\hat{\bf  q}\times\hat{\bf q}'+
{\cal A}^{ij}_{\rm tensor}S(\hat{\bf q})+\nonumber\\{\cal  A}^{ij}_{\rm exch.\
  tensor}S(\hat{\bf q}')+{\cal A}^{ij}_{\rm cm\  tensor}S(\hat{\bf P})+\nonumber\\
{\cal A}^{ij}_{\rm diff.\
  vector}i(\ve{\sigma}-\ve{\sigma}')\cdot\hat{\bf
  q}\times\hat{\bf P}+\nonumber\\{\cal A}^{ij}_{\rm cross\ vector}
(\ve{\sigma}\times\ve{\sigma}')\cdot(\hat{\bf  q}'\times\hat{\bf P}) +\nonumber\\
{\cal A}^{ij}_{q'P\ \rm so}{\hat{\bf q}}'\cdot {\hat{\bf
  P}}i(\ve{\sigma}+\ve{\sigma}')\cdot\hat{\bf  q}\times\hat{\bf
  P}+\nonumber\\{\cal A}^{ij}_{q'P\ \rm diff.}{\hat{\bf q}}'\cdot {\hat{\bf
  P}}i(\ve{\sigma}-\ve{\sigma}')\cdot\hat{\bf q}\times\hat{\bf P}+\nonumber\\
{\cal A}^{ij}_{q'P\ \rm quadr.\ so}{\hat{\bf q}}'\cdot {\hat{\bf P}}
\left(\ve{\sigma}\cdot{\hat{\bf q}}'\ve{\sigma}'\cdot{\hat{\bf P}}
+\right.\nonumber \\\ve{\sigma}\cdot{\hat{\bf P}} \ve{\sigma}'\cdot{\hat{\bf
    q}}'-\frac{2}{3}\left.{\hat{\bf q}}'\cdot{\hat{\bf P}}
\ve{\sigma}\cdot\ve{\sigma}'\right),\nonumber\\
\label{Eq:qp-int}
\eea
where the last three terms are zero for $i=j$.

We now comment on which terms in the interaction are relevant in
specific cases.  For the density response, the total nucleon number
density $n_{\rm p} + n_{\rm n}$, satisfies a local conservation law,
from which it follows from Eq.\ (\ref{chitp4}) that $ \lim_{\omega \rightarrow
0} \lim_{q \rightarrow 0} {\rm Im} \chi^{\rm tp} = 0$.  Similar
arguments apply for the isospin response. For the total spin response,
$ \lim_{\omega \rightarrow 0} \lim_{q \rightarrow 0} {\rm Im}
\chi^{\rm tp}$ is nonzero, and all terms in the interaction apart from
the scalar and spin-exchange terms contribute.  For the spin-isospin
response, all terms except the scalar term contribute: for this case
the spin-exchange term contributes because, while it conserves spin,
it does not conserve spin-isospin.

\section{Concluding remarks}

Landau's theory of normal Fermi liquids is usually regarded as
containing only the effects of single-pair states.  In this paper we
have shown how the leading contributions to the two-pair response at
frequencies low compared with the Fermi energy and wave numbers small
compared with the Fermi wave number may be evaluated in terms of the
collision rate for two-quasiparticle scattering.  The treatment brings
out clearly the important role played by conservation laws.  Our main
result is Eq.\ (\ref{chitp4}). Our calculation brings out  that there
are two time scales in the response:  one is the energy of a single
pair, $~v_{\rm F}q$ and the other is the quasiparticle lifetime.  The
structure on the energy scale $v_{\rm F}q$ is due to screening
 and to the presence of the intermediate quasiparticle close to the energy
shell. 

An important result of our analysis is that, for an operator which
does not represent a conserved quantity, the matrix element to a
two-pair state with excitation energy $\omega$ varies as $1/\omega$.
Examples of this behavior in specific applications have previously
been found in work on pair bremsstrahlung Refs. \cite{frimax,schwenk1,raffelt2,
hanhart, vandalen} and neutrino scattering \cite{raffelt2}.
At nonzero temperature the  growth of the matrix element for $\omega \rightarrow 0$ will be cut off 
by the Landau-Pomeranchuk-Migdal (LPM) effect. Neglecting the LPM effect, the multipair contribution to Im$\chi$ at zero temperature varies as $\omega$ and atnonzero temperatures small compared with the Fermi energy by $T^2/\omega$. 

Since completing the above work, we became acquainted with Ref. \cite{margueron} in which two-pair contributions to response functions were calculated.  The calculations were performed for a central interaction, and therefore from the general arguments that we have made, at long wavelengths one would expect the contribution of multipair states to vanish identically.  In the calculations of Ref.\cite{margueron} the wavenumber of the response was nonzero, and the authors demonstrated that there is a large cancellation between self-energy corrections to the particle propagators and vertex corrections: this is due to the fact that their interaction conserved both particle number and total spin. 
The contributions to the response investigated in the present paper are the ones that dominate at long wavelengths, and they have their origin is completely different: they are du to the noncentral nature of the nucleon-nucleon interaction.

For applications to astrophysical situations, it is important to
extend the long-wavelength results to finite wavelengths and higher
frequencies.  Of particular importance for these applications is the
evaluation of the two-quasiparticle scattering amplitudes.  These have
recently been calculated by Schwenk and Friman \cite{friman} including all contributions to second order in the low-momentum effective interaction $V_{{\rm low} k}$ \cite{sfb,schwenk2},  and it is
desirable to include higher-order effects.  In the case of neutrino
pair bremsstrahlung, $\omega > cq$, and therefore $\omega/v_{\rm F}q >
c/v_{\rm F}$ which is significantly larger than unity.  Consequently,
it is a good approximation to take the limit $v_{\rm F}q/\omega =0$,
as has generally been done \cite{hanhart,vandalen,schwenk1}.  In the
case of neutrino scattering, effects on the energy scale $v_{\rm F}q$
could be important, since the frequencies of interest are less than
$cq$.

Another direction for future work is to extend the calculations to
operators that change the isospin.  Examples are charged-current weak
interactions, which enter in the rate of the modified Urca process.  

\begin{acknowledgments}
We are grateful to Achim Schwenk for useful discussions and helpful comments. One of us (EO)
acknowledges financial support in part from a European Commission Marie
Curie Training Site Fellowship under Contract No.  HPMT-2000-00100.
This work has been conducted within the framework of the school on
Advanced Instrumentation and Measurements (AIM) at Uppsala University
supported financially by the Foundation for Strategic Research (SSF).
One of us (GL) acknowledges NORDITA for a kind hospitality during this work.
 
\end{acknowledgments} 

\appendix*
\section{Vertex corrections}\label{appendix}

Here we calculate the vertex correction factors $\lambda$ defined in Eq.\ (\ref{lambdadef}).  In terms of Landau theory, the vertex correction is given by

\beq
\lambda_{\bf p} =\frac{\delta \epsilon_{\bf p}^i}{\delta
  U^j}=\delta_{ij}+\sum_{k}\sum_{\bf p'} f_{\bf p p'}^{ik}
\frac{\delta n_{\bf p'}^k}{\delta U^j},
\label{vertexcorr}
\eeq

In the limit of no collisions between quasiparticles, the 
linearized quantum kinetic equation reads \cite[p.21]{bp}:
\bea
[\omega -(\epsilon^{i}_{{\bf p}+{\bf q}/2}-\epsilon^{i}_{{\bf
    p}-{\bf q}/2})]\delta n^{i}_{\bf p}+\nonumber\\(n^{i}_{{\bf p}+{\bf q}/2}-n^{i}_{{\bf p}-{\bf q}/2})\delta \epsilon^{i}_{\bf p}=0
\label{kineticEQ}
\eea
where the deviation of the quasiparticle energy from equilibrium is given by
\beq
\delta \epsilon^{i}_{\bf p}=\delta U^i + \sum_j \sum_{\bf p'} f_{\bf p p'}^{ij}\delta n^{j}_{\bf p'}.
\label{deltaepsilon}
\eeq
Here $U^i$ is the external potential acting on species $i$, and $f_{\bf p p'}^{ij}$ is the Landau quasiparticle interaction. Since we
consider only systems which are not spin polarized in equilibrium, we shall,
for simplicity, suppress spin indices. The calculation of the spin-density response is completely equivalent to the calculation of the density response, apart from the replacement of spin-symmetric Landau parameters by the spin-antisymmetric ones. 
For definiteness, let us consider an external potential $U^{\rm n}$
which couples to the neutron number. Inserting Eq. (\ref{deltaepsilon}) into Eq. (\ref{kineticEQ}) gives
\bea
[\omega -(\epsilon^{i}_{{\bf p}+{\bf q}/2}-\epsilon^{i}_{{\bf
    p}-{\bf q}/2})]\delta n^{i}_{\bf p}+\nonumber\\(n^{i}_{{\bf p}+{\bf q}/2}-n^{i}_{{\bf p}-{\bf q}/2})(\delta U^i + \sum_j \sum_{\bf p'} f_{\bf p p'}^{ij}\delta n^{j}_{\bf p'})=0.\label{kineticEQ2}
\eea
We expand $\delta n^{i}_{\bf p}$ in Legendre polynomials and write
\beq
\delta n^{i}_{\bf p}=\frac{1}{N^{i}(0)}\frac{n^{i}_{{\bf p}+{\bf
      q}/2}-n^{i}_{{\bf p}-{\bf q}/2}}{\epsilon^{i}_{{\bf p}+{\bf
      q}/2}-\epsilon^{i}_{{\bf p}-{\bf q}/2}}\sum_{l}\delta n_l^{i} P_l (\hat{\bf p}\cdot\hat{\bf q}), \label{dnexpansion}
\eeq
where $N^i(0) =m^*_i p_{\rm F}^i/\pi^2$ is the density of states at
the Fermi surface. The normalization factor is chosen so that the
change in the total density of species $i$ is $\delta n^i_0$.  The
Landau quasiparticle interaction is also expanded in Legendre
polynomials
\beq
f_{\bf p p'}^{ij} = \sum_{l} f_l^{ij}P_l (\hat{\bf p}\cdot\hat{\bf p'}).\label{Landauparameters}
\eeq
We now solve Eq.\ (\ref{kineticEQ}).  Since tensor contributions to
the quasiparticle interaction are generally small compared with the
scalar and spin exchange terms \cite{friman, ohp}, we shall neglect
the tensor terms.  In addition, we shall take into account only the
$l=0$ and $l=1$ terms in the scalar and spin-exchange terms. By summing Eq. (\ref{kineticEQ}) for neutrons and for protons over momenta, we obtain the relations
\bea
\omega \delta n_0^{\rm n} = q v_{\rm F}^{\rm n}\left(1+\frac{F_1^{\rm
    nn}}{3}\right)\frac{\delta n_1^{\rm n}}{3}
+q v_{\rm F}^{\rm n} N^{\rm n}(0)\frac{f_1^{\rm pn}}{3}\frac{\delta n_1^{\rm p}}{3}
\label{conn}
\eea
and
\bea
\omega \delta n_0^{\rm p} = q v_{\rm F}^{\rm p}\left(1+\frac{F_1^{\rm
    pp}}{3}\right)\frac{\delta n_1^{\rm p}}{3}+
q v_{\rm F}^{\rm p} N^{\rm p}(0)\frac{f_1^{\rm pn}}{3}\frac{\delta n_1^{\rm n}}{3}.
\label{conp}
\eea
These equations are an expression of the conservation laws for neutron number and proton number.
From Eqs.\ (\ref{conn}) and (\ref{conp}), we find 
\bea
\frac{\delta n_1^{\rm n}}{3} =\left[\frac{\omega}{q v_{\rm F}^{\rm n}}\left(1+\frac{F_1^{\rm pp}}{3}\right)\delta n_0^{\rm n}-\frac{\omega}{q v_{\rm F}^{\rm p}}\frac{N^{\rm n}(0) f_1^{\rm pn}}{3}\delta n_0^{\rm p}\right]
\nonumber\\\times\frac{1}{1+F_1^{\rm pp}/3+F_1^{\rm nn}/3+[F_1^{\rm pp}F_1^{\rm
    nn}-(F_1^{\rm pn})^2]/9}\nonumber\\\label{nn1}
\eea
and
\bea
\frac{\delta n_1^{\rm p}}{3} =\left[\frac{\omega}{q v_{\rm F}^{\rm p}}\left(1+\frac{F_1^{\rm nn}}{3}\right)\delta n_0^{\rm p}-\frac{\omega}{q v_{\rm F}^{\rm n}}\frac{N^{\rm p}(0)f_1^{\rm pn}}{3}\delta n_0^{\rm n}\right]
\nonumber\\\times\frac{1}{1+F_1^{\rm pp}/3+F_1^{\rm nn}/3+[F_1^{\rm pp}F_1^{\rm
      nn}-(F_1^{\rm pn})^2]/9}\nonumber\\\label{np1}
\eea
where 
\beq
F_l^{ij}=\sqrt{N^{i}(0)N^{j}(0)}f_l^{ij}.
\eeq
Rewriting Eq. (\ref{kineticEQ2}) as
\beq
\delta n^{i}_{\bf p}=\chi_{\bf p}^{i}(-\delta U^i - \sum_j \sum_{\bf p'} f_{\bf p p'}^{ij}\delta n^{j}_{\bf p'}),
\eeq
with
\beq
\chi_{\bf p}^{i}=\frac{(n^{i}_{{\bf p}+{\bf q}/2}-n^{i}_{{\bf p}-{\bf q}/2})}{(\omega -(\epsilon^{i}_{{\bf p}+{\bf q}/2}-\epsilon^{i}_{{\bf
    p}-{\bf q}/2}))},
\eeq
expanding $\delta n^{i}_{\bf p}$ according to Eq. (\ref{dnexpansion}) and summing over momenta,
we find two more relations for $\delta n_0^{\rm n}$, $\delta n_0^{\rm p}$, $\delta n_1^{\rm n}$ and $\delta n_1^{\rm p}$:
\bea
\delta n_0^{\rm n}=\chi_0^{\rm n}(-\delta U^{\rm n} -[F_0^{\rm nn}\delta
n_0^{\rm n}+\frac{F_1^{\rm nn}}{3}\frac{\omega}{qv_{\rm F}^{\rm n}}\delta n_1^{\rm n}]-\nonumber\\N^{\rm
  n}(0)[f_0^{\rm pn}\delta
n_0^{\rm p}+\frac{f_1^{\rm pn}}{3}\frac{\omega}{qv_{\rm F}^{\rm n}}\delta n_1^{\rm
  p}])\label{nn0}
\eea
and
\bea
\delta n_0^{\rm p}=\chi_0^{\rm p}(-[F_0^{\rm pp}\delta n_0^{\rm
  p}+\frac{F_1^{\rm pp}}{3}\frac{\omega}{qv_{\rm F}^{\rm p}}\delta n_1^{\rm p}]-\nonumber\\N^{\rm
  p}(0)[f_0^{\rm pn}\delta
n_0^{\rm n}+\frac{f_1^{\rm pn}}{3}\frac{\omega}{qv_{\rm F}^{\rm p}}\delta n_1^{\rm n}]),\label{np0}
\eea
where $\chi_0^i$ is 
\bea
\chi_0^i & =& \sum_{\bf p} \chi_{\bf p}^{i}\nonumber\\
&=& N^i(0)\int_{-1}^1 \frac{d\mu}{2} \frac{\mu}{\mu - (\omega/v_{\rm F}^iq)},
\eea
with $\mu = \hat{\bf p}\cdot\hat{\bf q}$.
Inserting Eqs. (\ref{nn0}) and (\ref{np0}) into Eqs. (\ref{nn1}) and
(\ref{np1}) gives for the response functions
\bea
\frac{\delta n_0^{\rm n}}{\delta U^{\rm n}}&=&-\chi_0^{\rm n}(1+\alpha^{\rm p}
  \chi_0^{\rm p})\times\nonumber\\
&&\frac{1}{1+\alpha^{\rm p} \chi_0^{\rm p}+  \alpha^{\rm n}
  \chi_0^{\rm n}+\alpha^{\rm p}
  \chi_0^{\rm p} \alpha^{\rm n} \chi_0^{\rm n}+\chi_0^{\rm
    n}\beta^{\rm n}\beta^{\rm p}\chi_0^{\rm p}}\nonumber\\
\label{deltann}
\eea
and
\bea
\frac{\delta n_0^{\rm p}}{\delta U^{\rm n}}&=&\chi_0^{\rm p}\beta^{\rm p}\chi_0^{\rm
    n}\times\nonumber\\&&\frac{1}{1+\alpha^{\rm p} \chi_0^{\rm p}+  \alpha^{\rm n}
  \chi_0^{\rm n}+\alpha^{\rm p} \chi_0^{\rm p} \alpha^{\rm n}
  \chi_0^{\rm n}+\chi_0^{\rm n}\beta^{\rm n}\beta^{\rm p}\chi_0^{\rm p}},\nonumber\\\label{deltanp}
\eea
where
\bea
\alpha^{\rm p} &=& F_0^{\rm pp} +\nonumber\\ 
\left(\frac{\omega}{q v_{\rm F}^{\rm p}}\right)^2\!\! &\times&\!\!\frac{F_1^{\rm pp}+[F_1^{\rm pp}F_1^{\rm
    nn}-(F_1^{\rm pn})^2]/3}{1+F_1^{\rm pp}/3+F_1^{\rm nn}/3+[F_1^{\rm
    pp}F_1^{\rm nn}-(F_1^{\rm pn})^2]/9}\nonumber\\\label{alphap}
\eea
and
\bea
\beta^{\rm p} &=&N^{\rm p}(0)\left[\frac{}{} f_0^{\rm pn}+\right. \nonumber\\\left(\frac{\omega}{q
      }\right)^2\!\!\!\frac{f_1^{\rm pn}}{v_{\rm F}^{\rm p} v_{\rm
        F}^{\rm n}}\!\!\!&\times&\!\!\!\left.\frac{1}{1+F_1^{\rm pp}/3+F_1^{\rm nn}/3+[F_1^{\rm
        pp}F_1^{\rm nn}-(F_1^{\rm pn})^2]/9}\right]\!\!.\nonumber\\\label{betap}
\eea 
The corresponding results for $\alpha^{\rm n}$ and $\beta^{\rm n}$
are obtained by interchanging p and n in Eqs. (\ref{alphap}) and
(\ref{betap}). If we instead use an external potential $\delta U^{\rm
  p}$ that couples to the proton number we get analogously,
\bea
\frac{\delta n^{\rm n}}{\delta U^{\rm p}}&=&\nonumber\\&&\frac{\chi_0^{\rm
    n}\beta^{\rm n}\chi_0^{\rm p}}{1+\alpha^{\rm p} \chi_0^{\rm p}+
  \alpha^{\rm n} \chi_0^{\rm n}+\alpha^{\rm p} \chi_0^{\rm p}
  \alpha^{\rm n} \chi_0^{\rm n}+\chi_0^{\rm n}\beta^{\rm n}\beta^{\rm
    p}\chi_0^{\rm p}}\nonumber\\
\eea
and
\bea
\frac{\delta n^{\rm p}}{\delta U^{\rm p}}&=&\nonumber\\&&\frac{-\chi_0^{\rm p}(1+\alpha^{\rm n}\chi_0^{\rm n})}{1+\alpha^{\rm p} \chi_0^{\rm p}+  \alpha^{\rm n} \chi_0^{\rm n}+\alpha^{\rm p} \chi_0^{\rm p} \alpha^{\rm n} \chi_0^{\rm n}+\chi_0^{\rm n}\beta^{\rm n}\beta^{\rm p}\chi_0^{\rm p}}.\nonumber\\
\eea

\noindent
For a one component system one finds  
\bea
\frac{\delta n_0}{\delta U} = \frac{-\chi_0 N(0)}{1+F_0\chi_0+(\omega/v_{\rm F}q)^2
F_1 \chi_0/(1+F_1/3)}
\eea
and
\bea 
\delta n_1 = \frac{\omega}{v_{\rm F}q}\frac{3}{1+F_1/3}\delta n_0,
\eea
We can now find an expression for the vertex corrections from Eq. (\ref{vertexcorr}).
Expanding $\delta n_{\bf p'}$ and $f_{\bf p p'}^{ik}$ in Legendre
polynomials, (Eq. (\ref{dnexpansion}) and Eq. (\ref{Landauparameters})), and keeping only the $l=0$ and $l=1$ terms gives
\beq
\sum_{\bf p'} f_{\bf p p'}^{ik} \frac{\delta n_{\bf p'}^k}{\delta
  U^j}=f_0^{ik} \frac{\delta n_0^k}{\delta U^j}+(\hat{\bf p}\cdot\hat{\bf q})\frac{f_1^{ik}}{3}\frac{\delta n_1^k}{\delta U^j}
\eeq
and therefore
\bea
\frac{\delta \epsilon_{\bf p}^{\rm n}}{\delta U^{\rm n}}&=& 1 + \left[\frac{}{}f_0^{\rm np}\right.\nonumber\\
\!\!&+&\!\!\!\left.\frac{(\hat{\bf p}\cdot\hat{\bf q})\omega/(qv_{\rm F}^{\rm p})f_1^{\rm np}}{1+F_1^{\rm pp}/3+F_1^{\rm nn}/3+[F_1^{\rm pp}F_1^{\rm nn}-(F_1^{\rm pn})^2]/9}\right]\!\!\frac{\delta n_0^{\rm p}}{\delta U^{\rm n}} \nonumber\\
&+&\frac{1}{N^{\rm n}(0)}\left[\frac{}{}F_0^{\rm nn}\right.\nonumber\\
\!\!&+&\!\!\!\left.\frac{(\hat{\bf p}\cdot\hat{\bf q})\omega/(qv_{\rm F}^{\rm n})(F_1^{\rm nn}+[F_1^{\rm nn}F_1^{\rm pp}+(F_1^{\rm pn})^2]/3)}{1+F_1^{\rm pp}/3+F_1^{\rm nn}/3+[F_1^{\rm
        pp}F_1^{\rm nn}-(F_1^{\rm pn})^2]/9}\right]\!\!\frac{\delta
    n_0^{\rm n}}{\delta U^{\rm n}}\nonumber\\
\eea
and
\bea
\frac{\delta \epsilon_{\bf p}^{\rm p}}{\delta U^{\rm n}}&=&\left[\frac{}{}f_0^{\rm np}\right.\nonumber\\
\!\!\!&+&\!\!\!\left.\frac{(\hat{\bf p}\cdot\hat{\bf q})\omega/(qv_{\rm F}^{\rm n})f_1^{\rm np}}{1+F_1^{\rm pp}/3+F_1^{\rm nn}/3+[F_1^{\rm pp}F_1^{\rm nn}-(F_1^{\rm pn})^2]/9}\right]\!\!\frac{\delta n_0^{\rm n}}{\delta U^{\rm n}}\nonumber\\ \!\!\!&+&\!\!\!\frac{1}{N^{\rm p}(0)}\left[\frac{}{}F_0^{\rm pp}\right.\nonumber\\
\!\!\!&+&\!\!\!\left.\frac{(\hat{\bf p}\cdot\hat{\bf q})\omega/(qv_{\rm F}^{\rm n})(F_1^{\rm nn}+[F_1^{\rm nn}F_1^{\rm pp}+(F_1^{\rm pn})^2]/3)}{1+F_1^{\rm pp}/3+F_1^{\rm nn}/3+[F_1^{\rm pp}F_1^{\rm nn}-(F_1^{\rm pn})^2]/9}\right]\!\!\frac{\delta n_0^{\rm p}}{\delta U^{\rm n}}.\nonumber\\
\eea

For a one component system this reduces to
\beq
\frac{\delta \epsilon_{\bf p}}{\delta U}=1+f_0 \frac{\delta n_0}{\delta
  U}+(\hat{\bf p}\cdot\hat{\bf q})\frac{f_1}{3}\frac{\delta
  n_1}{\delta U},
\eeq
which gives
\bea
\frac{\delta \epsilon_{\bf p}}{\delta U}&=&\left(1-\left[\hat{\bf p}\cdot\hat{\bf q}-\frac{\omega}{v_{\rm F} q}\right]\frac{\omega}{v_{\rm F} q}\frac{F_1\chi_0}{1+F_1/3}\right)\nonumber\\&\times&\!\!\!\frac{1}{1+\left[F_0+(\omega /v_{\rm F} q)^2 F_1/(1+F_1/3)\right]\chi_0}.\nonumber\\
\eea
For $\omega \gg v_{\rm F}q$, the vertex functions reduce to $\delta \epsilon_i/\delta U_j =\delta_{ij}$.
The response to a field that couples to the nucleon spin is given by exactly the same expressions, but with the spin-independent Landau parameters $F_l$ replaced by the spin-dependent ones $G_l$.  We note that in the case of the spin-dependent response, the response functions calculated here correspond only to the quasiparticle (Landau) contribution, and do not include the part due to multipair intermediate states.


\begin{thebibliography}{99}
\bibitem{traces}  Here we work in a basis of spin states in which the $\lambda^i$ are diagonal.  In a general basis, sums over spin states must be replaced by traces.  For example, the the term involving $(\xi_1^{{\cal O}})^*\xi_2^{\cal O}$ that would appear in Eq.\ \ref{chitp} is
${\rm Tr}[(\xi_1^{{\cal O}})^*_{\alpha \alpha'}{\cal A}_{1\alpha', 2 \beta}^{3 \gamma, 4 \delta}(\xi_2^{\cal O})_{\beta \beta'} ({\cal A}^{\dagger})_{1 \alpha, 2\beta'}^{3 \gamma,4 \delta}]$, where all repeated indices are to be summed over and the numbers 1,\ldots 4 refer only to the quasiparticle momentum..
\end{thebibliography}
\end{document}